\documentclass[a4paper,11pt]{article}
\pdfoutput=1 

\usepackage{jheppub} 

\title{\boldmath Index Theory and Supersymmetry of 5D Horizons}

\author[a]{J. Grover,}
\author[b]{J. Gutowski,}
\author[c]{G. Papadopoulos}
\author[d]{and W. A. Sabra}
\affiliation[a]{Physics Department \\University of Aveiro and I3N\\
Campus de Santiago,
3810-183 Aveiro, Portugal}
\affiliation[b]{Department of Mathematics \\
University of Surrey \\
Guildford, GU2 7XH, UK}
\affiliation[c]{Department of Mathematics\\
King's College London\\
Strand\\
London WC2R 2LS, UK}
\affiliation[d]{Centre for Advanced Mathematical Sciences and
Physics Department, \\
American University of Beirut \\ Lebanon}
\emailAdd{jai@ua.pt}
\emailAdd{j.gutowski@surrey.ac.uk}
\emailAdd{george.papadopoulos@kcl.ac.uk}
\emailAdd{ws00@aub.edu.lb}

\abstract{We prove that the near-horizon geometries of minimal gauged five-dimensional
supergravity preserve at least half of the supersymmetry. If the near-horizon geometries preserve a larger fraction, then they are locally isometric
to $AdS_5$. Our proof is based
on  Lichnerowicz type theorems for two horizon Dirac operators
constructed from the supercovariant connection  restricted to the horizon sections, and on an application of the index theorem. An application is  that all half-supersymmetric five-dimensional horizons admit an $\mathfrak{sl}(2, {\mathbb{R}})$ symmetry subalgebra.}

\arxivnumber{1303.0853}

\begin{document}
\maketitle
\flushbottom

\section{Introduction}

It is well known that near horizon geometries typically preserve more supersymmetries than the original black hole solutions.
This has been demonstrated for many supersymmetric black holes and branes, see eg \cite{gibbons}, and it is believed that it may be a universal property
of black hole solutions, at least in the theories without higher curvature corrections.
This supersymmetry enhancement is instrumental in understanding the topology and geometry of black hole horizons as additional supersymmetries
will impose additional restrictions on the topology and geometry of horizon sections, and this may lead to new insights into higher dimensional supersymmetric black holes, see eg \cite{israel}-\cite{susyring}
for some historical and more recent references in the subject. Furthermore, supersymmetry enhancement at the horizons has applications in the investigation of  properties of black hole systems like the entropy microstate counting, see eg \cite{sen}, and in AdS/CFT \cite{maldacena}.
If near horizon geometries of black holes
exhibit supersymmetry enhancement, then they should preserve at least two supersymmetries.
Although there are many partial results, a general understanding of which black holes
exhibit such supersymmetry enhancement near the horizons is not yet available.

In this paper we shall demonstrate that under some smoothness assumptions\footnote{The smoothness assumptions are necessary as the near horizon geometry
 of the NS5-brane preserves the same number of supersymmetries as the NS5-brane and so there is no supersymmetry enhancement. But the NS5 brane exhibits a singular dilaton at the horizon.} the near horizon black hole geometries of minimal 5-dimensional gauged supergravity
preserve at least half of the supersymmetry. In addition, if the near horizon geometries preserve a larger fraction of supersymmetry, then
they are locally isometric to $AdS_5$ and the 2-form field strength $F$ vanishes.  Furthermore a similar
argument to that presented in detail in \cite{mhor} implies that all half-supersymmetric 5-dimensional gauged supergravity horizons admit an $\mathfrak{sl}(2,{\mathbb{R}})$ symmetry subalgebra\footnote{The $\mathfrak{sl}(2,{\mathbb{R}})$ symmetry
of 5-dimensional horizons has been explored from a different point of view in \cite{lucietti}.}.

Our proof is topological in nature and relies on the compactness of the horizon sections. The analysis begins with the identification of independent field equations and  Killing spinor equations\footnote{Unlike most previous investigations of near horizon geometries, however see \cite{4dhor}, we do not impose the bi-linear matching condition, i.e. we do not identify
 the stationary Killing vector field of a black hole with the vector Killing spinor bilinear.}(KSEs) after appropriately integrating along the lightcone directions. Next, the Killing spinors are related to the zero modes of two horizon Dirac operators which are constructed from the supercovariant derivative of the supergravity theory appropriately restricted on the horizon sections. This relation is demonstrated via the proof of Lichnerowicz type theorems for the two
 horizon Dirac operators, utilizing the compactness of horizon sections.
 After this, we count the number of supersymmetries preserved by the near horizon geometries using  the vanishing of the index of one of the two horizon Dirac operators, and establish our result. The index of the horizon Dirac operator vanishes because it  has the same principal symbol as the $U(1)$ twisted Dirac operator and it is defined on the horizon sections which are 3-dimensional manifolds \cite{atiyah1}.

Although several steps of our proof rely on details of minimal gauged 5-dimensional supergravity,
we believe that it is likely that odd-dimensional supergravity near horizon geometries  preserve at least two supersymmetries.  Supporting evidence for this comes
from a similar calculation for M-horizons which have been shown to preserve an even number of supersymmetries \cite{mhor}.

This paper is organized as follows. In section two, we describe the near horizon fields of minimal gauged 5-dimensional supergravity and establish the
independent field equations.  In section three, we integrate the KSEs along the lightcone directions and present the independent KSEs.  In section four,
we describe some geometric properties of the backgrounds. In section five, we prove the two Lichnerowicz type  theorems. In section six, we prove our result
using the vanishing of the index for the horizon Dirac operators.
In section seven, we examine the ${\mathfrak{sl}}(2,\mathbb{R})$ symmetry of the half-supersymmetric solutions, and in section eight we give our conclusions.

\section{Near-Horizon Geometry and Field Equations}

The near horizon geometries of black holes with an active 2-form field strength can be expressed in Gaussian Null co-ordinates
 \cite{isen, gnull} as
\begin{eqnarray}
ds^2 &=& 2 {\bf{e}}^+ {\bf{e}}^- + \delta_{ij} {\bf{e}}^i {\bf{e}}^j~,~~~
\cr
F&=&-{\sqrt{3} \over 2} \Phi\, {\bf{e}}^+\wedge {\bf{e}}^--{\sqrt{3} \over 2} r {\bf{e}}^+\wedge d_h \Phi+{1\over2} dB_{ij}\, {\bf{e}}^i\wedge {\bf{e}}^j
\end{eqnarray}
where $d_h\Phi=d\Phi-h \Phi$, and we have used the frame
\begin{eqnarray}
\label{nhbasis}
{\bf{e}}^+ &=& du~,~~~
{\bf{e}}^- = dr + r h - {1 \over 2} r^2 \Delta du~,~~~
{\bf{e}}^i = e^i{}_J dy^J~,
\end{eqnarray}
$i,j=1,2,3$,  $u,r$ are the lightcone coordinates, and $h, \Delta, \Phi, B$ and ${\bf{e}}^i$  depend only on the  coordinates $y^I$, $I=1, 2, 3$, transverse to the lightcone. The black hole stationary
Killing vector field is identified with $\partial_u$. The 1-form gauge potential associated to $F$ is
\begin{eqnarray}
A = {\sqrt{3} \over 2} r \Phi du + B~.
\end{eqnarray}
Our smoothness assumption asserts that $\Delta, \Phi$, $h$, and $dB$ are globally defined scalars, 1-form and a closed 2-form
on the horizon section ${\cal S}$ given by $r=u=0$. Clearly the induced metric on ${\cal S}$ is
\begin{eqnarray}
ds_{\cal{S}}^2 = \delta_{ij} {\bf{e}}^i {\bf{e}}^j
\end{eqnarray}
and ${\cal S}$ is taken to be compact, connected without boundary. We denote the Levi-Civita connection of ${\cal{S}}$ by ${{\hat{\nabla}}}$.

The bosonic action is \cite{gst}
\begin{eqnarray}
 \mathcal{S}=\frac{1}{4\pi G}\int \left(\frac{1}{4}(R+{12 \over \ell^2})\star
1-\frac{1}{2}F\wedge \star F-\frac{2}{3\sqrt{3}}F\wedge F\wedge A
\right) \ ,
\end{eqnarray}
$F=dA$ is a $U(1)$ field strength and $\ell$ is a real nonzero
constant, using the same conventions as in \cite{reallads}. The
equations of motion are
\begin{eqnarray}
\label{einsteq}
R_{\alpha\beta}-2F_{\alpha\gamma}F_{\beta}^{\
  \gamma}+\frac{1}{3}g_{\alpha\beta}(F^2+{12 \over \ell^2})=0 \ ,
\end{eqnarray}
and
\begin{eqnarray}
\label{gaugeq}
 d\star
F+\frac{2}{\sqrt{3}}F\wedge F=0 \ ,
\end{eqnarray}
where $F^2\equiv F_{\alpha\beta}F^{\alpha\beta}$.
The orientation is specified by
\begin{eqnarray}
\epsilon_5 = {\bf{e}}^+ \wedge {\bf{e}}^- \wedge \epsilon_3
\end{eqnarray}
where $\epsilon_5$ is the 5-dimensional volume form, and $\epsilon_3$ is the volume form on ${\cal{S}}$.
The Hodge dual on ${\cal{S}}$ is denoted by $\star_3$.

Before proceeding with the analysis of the supersymmetry, we consider the bosonic field equations.
From the gauge field equations one obtains the conditions:
\begin{eqnarray}
\label{geq1}
d \star_3 dB + {\sqrt{3} \over 2} \star_3 d_h \Phi  - h \wedge \star_3 dB -2 \Phi dB =0
\end{eqnarray}
and
\begin{eqnarray}
\label{geq2}
-dh \wedge \star_3 dB - {\sqrt{3} \over 2} h \wedge \star_3 d_h \Phi
+{\sqrt{3} \over 2} d \star_3 d_h \Phi  -2 d_h \Phi  \wedge dB =0
\end{eqnarray}
however we remark that ({\ref{geq1}}) implies ({\ref{geq2}}). In components ({\ref{geq1}}) and ({\ref{geq2}})
are equivalent to
\begin{eqnarray}
\label{geq3}
{{\hat{\nabla}}}^m (dB)_{mi} +(dB)_{im} h^m +2 \Phi (\star_3 dB)_i - {\sqrt{3} \over 2} (d_h\Phi)_i  =0
\end{eqnarray}
and
\begin{eqnarray}
\label{geq4}
-{1 \over 2} dh_{mn} dB^{mn} - {\sqrt{3} \over 2} h^i (d_h \Phi)_i
+{\sqrt{3} \over 2} {{\hat{\nabla}}}^i (d_h \Phi)_i - \epsilon^{ijk} (d_h \Phi)_i dB_{jk} =0 \ .
\nonumber \\
\end{eqnarray}

Next we consider the Einstein field equations. The $+-$ and $ij$ components of the
Einstein equations are
\begin{eqnarray}
\label{epm}
{1 \over 2} {{\hat{\nabla}}}^i h_i - \Delta - {1 \over 2} h^2 + \Phi^2 +{1 \over 3} dB_{mn} dB^{mn} +{4 \over \ell^2} =0
\end{eqnarray}
and
\begin{eqnarray}
\label{eij}
{\hat{R}}_{ij} = - {{\hat{\nabla}}}_{(i} h_{j)} +{1 \over 2} h_i h_j +2 dB_{im} dB_j{}^m -{1 \over 3} \delta_{ij} \big(-{3 \over 2} \Phi^2
+ dB_{mn} dB^{mn} +{12 \over \ell^2} \big)
\end{eqnarray}
respectively,
where ${\hat{R}}_{ij}$ denotes the Ricci tensor of ${\cal{S}}$. In addition, the $+i$ and $++$ components of the Einstein equations are
\begin{eqnarray}
\label{ei}
{1 \over 2} {{\hat{\nabla}}}^j dh_{ij}  - dh_{ij} h^j - {{\hat{\nabla}}}_i \Delta + \Delta h_i +{3 \over 2} \Phi (d_h \Phi)_i
+ \sqrt{3} (d_h \Phi)_j dB_i{}^j =0
\end{eqnarray}
and
\begin{eqnarray}
\label{epp}
{1 \over 2} {{\hat{\nabla}}}^2 \Delta - {3 \over 2} h^i {{\hat{\nabla}}}_i \Delta -{1 \over 2} \Delta {{\hat{\nabla}}}^i h_i
+ \Delta h^2 +{1 \over 4} dh_{ij} dh^{ij} -{3 \over 2}(d_h \Phi)_i(d_h \Phi)^i =0 \ .
\nonumber \\
\end{eqnarray}
However, the $+-$ and $ij$ components of the Einstein equations ({\ref{epm}}) and ({\ref{eij}}) together with
the gauge equations ({\ref{geq1}}) imply both ({\ref{ei}}) and ({\ref{epp}}); ({\ref{ei}}) is obtained by evaluating
the Bianchi identity associated with ({\ref{eij}}), and ({\ref{epp}}) is then found by taking the divergence of ({\ref{ei}}).
To summarize, the independent bosonic field equations are ({\ref{geq1}}), ({\ref{epm}}) and ({\ref{eij}}).

\section{Supersymmetric Near-Horizon Geometries}

\subsection{Integrability of the lightcone directions}
The KSE of minimal 5-dimensional gauged supergravity is
\begin{eqnarray}
\label{grav}
 \bigg[\nabla_\mu  - {i \over
4\sqrt{3}} F^{\nu_1 \nu_2} \Gamma_\mu \Gamma_{\nu_1 \nu_2} +{3i \over
2\sqrt{3}} F_{\mu}{}^{\nu} \Gamma_{\nu} +
{2 \sqrt{3} \over \ell} ({1\over4\sqrt{3}}\Gamma_{\mu} + {i\over2}A_{\mu}) \bigg]
\epsilon =0 \ .
\nonumber \\
\end{eqnarray}
where $\nabla$ is the Levi-Civita connection of spacetime and  $\epsilon$ is a Dirac spinor.
The representation of $\mathrm{Cliff}(4,1)$ used is specified in Appendix B, along with
other conventions, including the decomposition of $\epsilon = \epsilon_+ + \epsilon_-$
where $\epsilon_+, \epsilon_-$ are chiral spinors with respect to $\Gamma_{+-}$. Observe that the KSE is linear over the complex numbers. So the
supersymmetric configurations always admit even number of supersymmetries as counted over the real numbers.

We shall first integrate  ({\ref{grav}}) along the lightcone directions  $r$ and $u$.   Then
we shall establish the independent KSEs on the horizon section ${\cal S}$. For this, we shall make an extensive use of the
bosonic field equations listed in the previous section where appropriate.

To begin, consider the $\mu=-$ component of ({\ref{grav}}), this can be integrated to obtain:
\begin{eqnarray}
\label{rdec}
\epsilon_+ &=& \phi_+
\nonumber \\
\epsilon_- &=& r \Gamma_- \big( ({1 \over 4} h +{1 \over 2 \sqrt{3}} \star_3 dB)_i \Gamma^i -{i \over 2} \Phi -{1 \over 2 \ell}
\big) \phi_+ + \phi_-
\end{eqnarray}
where
\begin{eqnarray}
\partial_r \phi_\pm =0 \ .
\end{eqnarray}
Next we consider the $\mu=+$ component of ({\ref{grav}}). On evaluating this component at $r=0$, one obtains
\begin{eqnarray}
\label{phexp}
\phi_+ &=& u \Gamma_+ \big( ({1 \over 4} h -{1 \over 2 \sqrt{3}} \star_3 dB)_i \Gamma^i +{i \over 2} \Phi -{1 \over 2 \ell}
\big) \eta_- + \eta_+
\nonumber \\
\phi_- &=& \eta_-
\end{eqnarray}
where
\begin{eqnarray}
\partial_r \eta_{\pm} = \partial_u \eta_\pm =0 \ .
\end{eqnarray}
The remaining content of the $\mu=+$ component can be written as
\begin{eqnarray}
\label{alg1}
\bigg( 2 \big( ({1 \over 4} h -{1 \over 2 \sqrt{3}} \star_3 dB)_i \Gamma^i -{i \over 2} \Phi +{1 \over 2 \ell} \big)
 \big( ({1 \over 4} h +{1 \over 2 \sqrt{3}} \star_3 dB)_j \Gamma^j -{i \over 2} \Phi -{1 \over 2 \ell} \big)
\nonumber \\
+  {1 \over 2} \Delta +{3i \over 2 \ell} \Phi + \big({i \over 4} \star_3 dh_i -{i \over 4}  (d_h \Phi)_i  \big)
\Gamma^i
  \bigg) \phi_+ =0
\end{eqnarray}
 and
 \begin{eqnarray}
\label{alg2}
\bigg( 2 \big( -({1 \over 4} h +{1 \over 2 \sqrt{3}} \star_3 dB)_i \Gamma^i -{i \over 2} \Phi -{1 \over 2 \ell} \big)
 \big( ({1 \over 4} h -{1 \over 2 \sqrt{3}} \star_3 dB)_j \Gamma^j +{i \over 2} \Phi -{1 \over 2 \ell} \big)
\nonumber \\
-  {1 \over 2} \Delta +{3i \over 2 \ell} \Phi + \big(-{i \over 4} \star_3 dh_i -{3i \over 4} (d_h \Phi)_i  \big)
\Gamma^i
  \bigg) \phi_- =0
\end{eqnarray}
and
\begin{eqnarray}
\label{alg3}
\bigg( \big({3i \over 2 \ell} \Phi +  \big({i \over 4} \star_3 dh_i +{3i \over 4}  (d_h \Phi)_i  \big)
\Gamma^i \big) \big( ({1 \over 4} h +{1 \over 2 \sqrt{3}} \star_3 dB)_j \Gamma^j -{i \over 2} \Phi -{1 \over 2 \ell} \big)
\nonumber \\
+{1 \over 4} (\Delta h_i -{{\hat{\nabla}}}_i \Delta) \Gamma^i \bigg) \phi_+ =0 \ .
\end{eqnarray}
Next, consider the $\mu=i$ component of ({\ref{grav}}). Evaluating this component at $r=0$ one obtains
\begin{eqnarray}
\label{gv1}
{{\hat{\nabla}}}_i \phi_+ + \bigg( -{1 \over 4} h_i -{i \over 4} \Phi \Gamma_i
-{1 \over 2 \sqrt{3}} (\star_3 dB)_i +{i \over \sqrt{3}} dB_{ij} \Gamma^j
+{\sqrt{3}i \over \ell} B_i +{1 \over 2 \ell} \Gamma_i \bigg) \phi_+=0
\end{eqnarray}
and
\begin{eqnarray}
\label{gv2}
{{\hat{\nabla}}}_i \phi_- + \bigg( {1 \over 4} h_i +{i \over 4} \Phi \Gamma_i
+{1 \over 2 \sqrt{3}} (\star_3 dB)_i +{i \over \sqrt{3}} dB_{ij} \Gamma^j
+{\sqrt{3}i \over \ell} B_i +{1 \over 2 \ell} \Gamma_i \bigg) \phi_-=0
\end{eqnarray}
and the remaining content of the $\mu=i$ component of ({\ref{grav}}) is
\begin{eqnarray}
\label{alg4}
{{\hat{\nabla}}}_i \bigg( \big( ({1 \over 4} h +{1 \over 2 \sqrt{3}} \star_3 dB)_j \Gamma^j -{i \over 2} \Phi -{1 \over 2 \ell} \big)
\phi_+ \bigg)
\nonumber \\
+ \bigg(-{3 \over 4} h_i -{i \over 4} \Phi \Gamma_i +{1 \over 2 \sqrt{3}} (\star_3 dB)_i
-{i \over \sqrt{3}} dB_{ij} \Gamma^j +{\sqrt{3} i \over \ell} B_i -{1 \over 2 \ell} \Gamma_i \bigg)
\nonumber \\
\times \bigg(  \big({1 \over 4} h +{1 \over 2 \sqrt{3}} \star_3 dB)_k \Gamma^k -{i \over 2} \Phi -{1 \over 2 \ell}
\bigg) \phi_+
\nonumber \\
+ \bigg( -{1 \over 4} dh_{ij} \Gamma^j -{i \over 4} \Gamma_i  (d_h \Phi)_j \Gamma^j +{3i \over 4}
({{\hat{\nabla}}}_i \Phi - \Phi h_i) \bigg) \phi_+ =0 \ .
\end{eqnarray}
This concludes the analysis of the integrability of the KSEs along the lightcone directions.

\subsection{The KSEs ({\ref{alg1}}), ({\ref{alg2}}) and ({\ref{alg4}}) are not independent}

To find the supersymmetric solutions of supergravity theories, it is customary to first solve all the KSEs and then impose the field equations which are not implied as integrability
conditions of the KSEs. However, here we shall adopt a different strategy. We shall use the field equations to identify the independent KSEs on the horizon section ${\cal S}$.
To proceed, note that ({\ref{gv1}}) implies that
\begin{eqnarray}
\label{auxx1}
{1 \over 2} {\hat{R}}_{jk} \Gamma^k \phi_+ &=& \Gamma^i ({{\hat{\nabla}}}_i {{\hat{\nabla}}}_j - {{\hat{\nabla}}}_j {{\hat{\nabla}}}_i) \phi_+
\nonumber \\
&=& \bigg( \Gamma^i \big( {1 \over 4} dh_{ij} +{1 \over 2 \sqrt{3}}({{\hat{\nabla}}}_i \star_3 dB_j + {{\hat{\nabla}}}_j \star_3 dB_i)
-{\sqrt{3} i \over \ell} dB_{ij} \big)
\nonumber \\
&+&{i \over 4} {{\hat{\nabla}}}_i \Phi \Gamma^i{}_j - {i \over 2} {{\hat{\nabla}}}_j \Phi +{i \over \sqrt{3}} {{\hat{\nabla}}}^i dB_{ij}
-4 \big( {i \over 4} \Phi -{1 \over 2 \ell} \big)^2 \Gamma_j
\nonumber \\
&+& \big( {i \over 4} \Phi -{1 \over 2 \ell} \big) \big( {4 \over \sqrt{3}} \star_3 dB_j +{2i \over \sqrt{3}} dB_{jk} \Gamma^k \big)
+{2 \over 3} dB_j{}^i dB_{ki} \Gamma^k \bigg) \phi_+ \ .
\end{eqnarray}

On contracting ({\ref{auxx1}}) with $\Gamma^j$ and using ({\ref{eij}}) and ({\ref{epm}}) to rewrite the Ricci scalar of ${\cal{S}}$ in terms of
$\Delta$, one obtains after making use of ({\ref{geq1}}), the condition ({\ref{alg1}}). Hence we find that ({\ref{alg1}}) is implied
by the bosonic field equations and ({\ref{gv1}}).

Similarly, we find that ({\ref{gv2}}) implies that
\begin{eqnarray}
\label{auxx2}
{1 \over 2} {\hat{R}}_{jk} \Gamma^k \phi_- &=& \Gamma^i ({{\hat{\nabla}}}_i {{\hat{\nabla}}}_j - {{\hat{\nabla}}}_j {{\hat{\nabla}}}_i) \phi_-
\nonumber \\
&=& \bigg(- \Gamma^i \big( {1 \over 4} dh_{ij} +{1 \over 2 \sqrt{3}}({{\hat{\nabla}}}_i \star_3 dB_j + {{\hat{\nabla}}}_j \star_3 dB_i)
+{\sqrt{3} i \over \ell} dB_{ij}\big)
\nonumber \\
&+& {i \over 2} {{\hat{\nabla}}}_j \Phi +{i \over 4} {{\hat{\nabla}}}_i \Phi \Gamma_j{}^i +{i \over \sqrt{3}} {{\hat{\nabla}}}^i dB_{ij}
-4 \big( {i \over 4} \Phi +{1 \over 2 \ell} \big)^2 \Gamma_j
\nonumber \\
&+& \big( {i \over 4} \Phi +{1 \over 2 \ell} \big) \big( {4 \over \sqrt{3}} \star_3 dB_j -{2i \over \sqrt{3}} dB_{jk} \Gamma^k \big)
+{2 \over 3} dB_j{}^i dB_{ki} \Gamma^k \bigg) \phi_- \ .
\end{eqnarray}
On contracting ({\ref{auxx2}}) with $\Gamma^j$ and using ({\ref{eij}}) and ({\ref{epm}}) to
rewrite the Ricci scalar of ${\cal{S}}$ in terms of $\Delta$, one obtains after making use of ({\ref{geq1}})
the condition ({\ref{alg2}}). Hence the condition ({\ref{alg2}}) is implied by the bosonic field equations together with ({\ref{gv2}}). In addition, it is straightforward to see that the $u$-dependent part of ({\ref{gv1}})
(as we recall that $\phi_+$ contains a term linear in $u$ as given in ({\ref{phexp}})),
is in fact equivalent to ({\ref{auxx2}}). This follows on substituting ({\ref{gv1}}) and ({\ref{geq1}}) into
the $u$-dependent part of ({\ref{gv1}}).

Next consider ({\ref{alg4}}). This condition can be rewritten, using ({\ref{gv1}}), as:
\begin{eqnarray}
\bigg({1 \over 2 \sqrt{3}} {{\hat{\nabla}}}_i \star_3 dB_j \Gamma^j -{1 \over 8} h_i h_j \Gamma^j
-{i \over 4} \Phi h_i +{i \over 8} \Phi \Gamma_i{}^j h_j -{i \over 4} \Gamma_i{}^j {{\hat{\nabla}}}_j \Phi
\nonumber \\
-{1 \over 4 \sqrt{3}} (h \wedge \star_3 dB)_{ij} \Gamma^j -{i \over 2 \sqrt{3}} dB_{ij} h^j
+{1 \over 12} dB_{mn} dB^{mn} \Gamma_i -{1 \over 6} dB_{im} dB_j{}^m \Gamma^j
\nonumber \\
+ \big(-{\sqrt{3} \over 4} \Phi +{i \over \sqrt{3} \ell} \big) dB_{ij} \Gamma^j
-{1 \over 2 \sqrt{3}}(i \Phi + {2 \over \ell}) \star_3 dB_i
+{1 \over \ell}({i \over 2} \Phi +{1 \over 2 \ell}) \Gamma_i +{1 \over 4} {{\hat{\nabla}}}_j h_i \Gamma^j \bigg) \phi_+=0 \ .
\nonumber \\
\end{eqnarray}
However, this condition is equivalent to ({\ref{auxx1}}) on making use of the
Einstein equations ({\ref{eij}}) and the gauge equation ({\ref{geq1}}). Hence we also find that ({\ref{alg4}})
is implied by the bosonic field equations and ({\ref{gv1}}).

It remains to consider the condition ({\ref{alg3}}). We shall show in the remaining part of this section that
({\ref{alg3}}) is also implied by the bosonic field equations and ({\ref{gv1}}), although in order
to establish this, we shall make use of global
properties of ${\cal{S}}$.

\subsection{The KSE ({\ref{alg3}}) is not independent}

\subsubsection{Maximum principle}

To proceed with the analysis of the conditions on the spinors
imposed by the compactness of ${\cal{S}}$, we shall
assume that the Killing spinor is sufficiently regular
so that all gauge-invariant spinor
bilinears constructed from $\phi_\pm$  are smooth forms on ${\cal{S}}$.

It is useful to compute, using ({\ref{gv1}})
\begin{eqnarray}
\label{sder}
{{\hat{\nabla}}}_i \langle \phi_+, \phi_+ \rangle = \langle \phi_+ , \big( {1 \over 2} h_i +{1 \over \sqrt{3}} \star_3 dB_i -{1 \over \ell} \Gamma_i \big) \phi_+ \rangle \ .
\end{eqnarray}
Next ({\ref{epm}}) implies  that
\begin{eqnarray}
\label{lapl1}
{{\hat{\nabla}}}^i {{\hat{\nabla}}}_i  \langle \phi_+, \phi_+ \rangle
- \big( h^i -{2 \over \sqrt{3}} \star_3 dB^i \big) {{\hat{\nabla}}}_i  \langle \phi_+, \phi_+ \rangle
\nonumber \\
= \bigg( \Delta -{1 \over \ell^2} - \Phi^2 +{1 \over 4} h^2 +{1 \over \sqrt{3}} h^i \star_3 dB_i
+{1 \over 3} (\star_3 dB)^2 \bigg)  \langle \phi_+, \phi_+ \rangle~.
\end{eqnarray}

Moreover, ({\ref{alg1}}) (which we recall follows from ({\ref{gv1}}) together with the bosonic conditions),
implies that
\begin{eqnarray}
\label{lapl1b}
\langle \phi_+, \Delta \phi_+ \rangle = \langle \phi_+, \big(-{1 \over 4} h^2 +{1 \over 3} (\star_3 dB)^2
+{1 \over \ell^2} + \Phi^2 -{2 \over \sqrt{3} \ell} \star_3 dB_i \Gamma^i \big) \phi_+ \rangle~,
\end{eqnarray}
where to obtain this identity, we have taken the real part of the inner product of ({\ref{alg1}})
with $\phi_+$. On combining ({\ref{sder}}), ({\ref{lapl1}}) and ({\ref{lapl1b}}), one then obtains
\begin{eqnarray}
\label{max1}
{{\hat{\nabla}}}^i {{\hat{\nabla}}}_i \langle \phi_+,  \phi_+ \rangle - h^i {{\hat{\nabla}}}_i \langle \phi_+,  \phi_+ \rangle =0~,
\end{eqnarray}
and hence an application of the maximum principle implies that
\begin{eqnarray}
\langle \phi_+, \phi_+ \rangle = {\cal{F}}(u)~,
\end{eqnarray}
where ${\cal{F}}$ is a quadratic in $u$ with constant coefficients.

A similar argument, using ({\ref{gv2}}) and ({\ref{alg2}}) yields the condition
\begin{eqnarray}
\label{lapl2b}
{{\hat{\nabla}}}^i {{\hat{\nabla}}}_i  \langle \phi_-, \phi_- \rangle
+ \big( h^i -{4 \over \sqrt{3}} \star_3 dB^i \big) {{\hat{\nabla}}}_i  \langle \phi_-, \phi_- \rangle
\nonumber \\
= \bigg( -{1 \over 2} h^2 +2 \star_3 dB_i \star_3 dB^i +{2 \over \sqrt{3}} h^i \star_3 dB_i +{6 \over \ell^2}
\bigg)  \langle \phi_-, \phi_- \rangle \ .
\end{eqnarray}
It follows from Lemma 2 of {\cite{susyring}} that if $\phi_-$ vanishes at any point then
$\phi_- =0$ everywhere on ${\cal{S}}$.

\subsubsection{Solutions with $\phi_+=0$ everywhere on ${\cal{S}}$}

To proceed, first consider the special case for which $\phi_+=0$ everywhere on ${\cal{S}}$.
Then ({\ref{phexp}}) implies that
\begin{eqnarray}
\label{auxx3}
\bigg( ({1 \over 4} h -{1 \over 2 \sqrt{3}} \star_3 dB)_i \Gamma^i +{i \over 2} \Phi -{1 \over 2 \ell} \bigg)
\phi_-=0
\end{eqnarray}
and on substituting this into ({\ref{alg2}}) one further obtains
\begin{eqnarray}
\label{auxx3b}
\bigg( -{1 \over 2} \Delta +{3i \over 2 \ell} \Phi -{i \over 4} \star_3 dh_i \Gamma^i
-{3i \over 4} (d_h\Phi)_i \Gamma^i \bigg) \phi_-=0
\end{eqnarray}
and hence
\begin{eqnarray}
\Delta   \langle \phi_-, \phi_- \rangle=0 \ .
\end{eqnarray}
However, recall that if $\phi_+=0$ everywhere, then $\phi_-$ can never vanish anywhere,
for if it did, then both $\phi_+=0$ and $\phi_-=0$ everywhere, which implies the Killing spinor
vanishes everywhere. We discard this case.
Hence it follows that
\begin{eqnarray}
\Delta=0 \ .
\end{eqnarray}
Next consider ({\ref{auxx3}}) which implies that
\begin{eqnarray}
\langle \phi_- , \big( {1 \over 2} h_i -{1 \over \sqrt{3}} \star_3 dB_i -{1 \over \ell} \Gamma_i \big) \phi_-
\rangle =0~,
\end{eqnarray}
and  together with ({\ref{gv2}}) give that
\begin{eqnarray}
{{\hat{\nabla}}}_i  \langle \phi_-, \phi_- \rangle = - h_i  \langle \phi_-, \phi_- \rangle \ .
\end{eqnarray}
As $\phi_-$ is nowhere vanishing, this implies that
\begin{eqnarray}
dh=0
\end{eqnarray}
and the Einstein equation ({\ref{epp}}) further implies that
\begin{eqnarray}
d_h \Phi  =0
\end{eqnarray}
and then ({\ref{auxx3b}}) implies that
\begin{eqnarray}
\Phi=0
\end{eqnarray}
as well.
Furthermore, on substituting all these conditions back into the Einstein equation ({\ref{epm}})
one finds
\begin{eqnarray}
{{\hat{\nabla}}}^i {{\hat{\nabla}}}_i  \langle \phi_-, \phi_- \rangle = \bigg( {2 \over 3} dB_{mn} dB^{mn} +{8 \over \ell^2} \bigg)
 \langle \phi_-, \phi_- \rangle \ .
\end{eqnarray}
This leads to a contradiction, because the integral of the LHS over ${\cal{S}}$ vanishes,
whereas the integral of the RHS is positive.
Hence it follows that there are no solutions for which $\phi_+=0$ everywhere on ${\cal{S}}$.

\subsubsection{Solutions for which $\phi_+ \not\equiv 0$}

Having established that there are no solutions with $\phi_+ \equiv 0$, we note that
as ${{\hat{\nabla}}}_i  \langle \phi_+, \phi_+ \rangle =0$, ({\ref{lapl1}}) implies
\begin{eqnarray}
\label{auxx4}
 \Delta -{1 \over \ell^2} - \Phi^2 +{1 \over 4} h^2 +{1 \over \sqrt{3}} h^i \star_3 dB_i
+{1 \over 3} (\star_3 dB)^2 =0
\end{eqnarray}
and ({\ref{sder}}) also implies that
\begin{eqnarray}
\label{auxx4b}
 \langle \phi_+, \phi_+ \rangle ({1 \over 2} h_i + {1 \over \sqrt{3}} \star_3 dB_i)= {1 \over \ell}
 \langle \phi_+, \Gamma_i \phi_+ \rangle \ .
 \end{eqnarray}
 On taking the norm of both sides of this expression, one finds
 \begin{eqnarray}
 \label{auxx5}
 \big({1 \over 2} h +{1 \over \sqrt{3}} \star_3 dB \big)^2 ={1 \over \ell^2}~,
 \end{eqnarray}
 and on substituting this condition back into ({\ref{auxx4}}) one also obtains
 \begin{eqnarray}
 \label{auxx6}
 \Delta = \Phi^2 \ .
 \end{eqnarray}

 To continue, we remark that ({\ref{auxx4b}}) is equivalent to
 \begin{eqnarray}
 \label{auxx7}
 \bigg( \big({1 \over 2} h +{1 \over \sqrt{3}} \star_3 dB \big)_i \Gamma^i -{1 \over \ell} \bigg) \phi_+=0 \ .
 \end{eqnarray}
The condition ({\ref{alg3}}) has not been used at any stage of this analysis.
 Furthermore, on substituting ({\ref{auxx7}}) into both ({\ref{alg1}}) and ({\ref{alg3}}) it is straightforward to see
 that ({\ref{alg3}}) is implied by ({\ref{alg1}}).
 It therefore follows that ({\ref{alg3}}) is implied by the bosonic field equations and ({\ref{gv1}}).

 \subsection{The independent Killing spinor and field equations}

We have proven that all of the algebraic conditions on $\phi_\pm$, i.e. ({\ref{alg1}}), ({\ref{alg2}}),
 ({\ref{alg3}}) and ({\ref{alg4}}) are implied by the bosonic field equations and the reduced on ${\cal S}$ gravitino KSEs
 ({\ref{gv1}}) and ({\ref{gv2}}). We have also
 proven that the $u$-dependent part of ({\ref{gv1}}) is implied by ({\ref{gv2}}) and the bosonic field equations. Therefore we have demonstrated
that the necessary and sufficient conditions for a near-horizon
 geometry to be  a supersymmetric solution of minimal gauged supergravity are  the identifications
 \begin{eqnarray}
 \label{ident1}
 \Delta= \Phi^2~,
 \end{eqnarray}
 and
 \begin{eqnarray}
 \label{ident2}
  \big({1 \over 2} h +{1 \over \sqrt{3}} \star_3 dB \big)^2 ={1 \over \ell^2}~.
  \end{eqnarray}
In addition,  the background has  to satisfy the bosonic field equations
 ({\ref{geq1}}), ({\ref{epm}}) and ({\ref{eij}}) together with the horizon section KSEs
 \begin{eqnarray}
 \label{ggv1}
 \nabla_i^\pm\eta_\pm\equiv {{\hat{\nabla}}}_i \eta_\pm +\Psi_i^\pm \eta_\pm=0 ~,
 \end{eqnarray}
 where
 \begin{eqnarray}
\Psi^{\pm}_i = \mp({1 \over 4} h_i +{1 \over 2 \sqrt{3}} \star_3 dB_i) + \big( \mp {i \over 4} \Phi +{1 \over 2 \ell} \big) \Gamma_i
+{i \over \sqrt{3}} dB_{ij} \Gamma^j +{\sqrt{3} i \over \ell} B_i~,
\end{eqnarray}
and the spinors $\eta_\pm$ are $u,r$-independent.
The Killing spinor $\epsilon$ is then constructed from $\eta_\pm$ using ({\ref{rdec}})
and ({\ref{phexp}}).  Therefore, the number of supersymmetries preserved by the near horizon geometries is equal to the number
of linearly independent $\nabla^\pm$-parallel spinors.
These conditions, together with ({\ref{epm}}), also imply that there are
no solutions with $h=0$ everywhere on ${\cal{S}}$.

We can take, without loss of
generality,
$\eta_+ \neq 0$. To see this, note that if $\eta_+=0$, then $\eta_- \neq 0$. Moreover, we have shown that the spinor
\begin{eqnarray}
\eta'_+ =  {\partial \phi_+ \over \partial u} = \Gamma_+ \bigg(({1 \over 4} h -{1 \over 2 \sqrt{3}}
\star_3 dB)_i \Gamma^i +{i \over 2} \Phi -{1 \over 2 \ell} \bigg) \eta_-~,
\end{eqnarray}
also satisfies $\nabla^+\eta_+'=0$ in  ({\ref{ggv1}}). Furthermore, we must take $\eta'_+ \neq 0$, because
if both $\eta'_+=0$ and $\eta_+=0$, then $\phi_+=0$, and we have proven that this leads to a contradiction.

Thus $\nabla^+\eta_+=0$ in ({\ref{ggv1}}) always admits a non-vanishing solution  and so we take $\eta_+ \neq 0$.  Without loss of generality, we set
\begin{eqnarray}
\langle \eta_+, \eta_+ \rangle =1 \ .
\end{eqnarray}

 \section{Conditions on the Geometry}

Before continuing to examine the number of supersymmetries preserved by near horizon geometries, we
briefly present the conditions imposed on the geometry of ${\cal{S}}$ from
the results obtained so far.
It will be convenient to define
\begin{eqnarray}
\label{zdef}
Z_i = \langle \eta_+, \Gamma_i \eta_+ \rangle \ .
\end{eqnarray}
It follows from ({\ref{auxx7}}) that
\begin{eqnarray}
\label{auxx8}
{1 \over 2} h +{1 \over \sqrt{3}} \star_3 dB = {1 \over \ell} Z~,
\end{eqnarray}
and note that
\begin{eqnarray}
\label{auxx8b}
Z^2=1 \ .
\end{eqnarray}
Then on taking the covariant derivative of $Z$ using ({\ref{ggv1}}),
one obtains
\begin{eqnarray}
\label{auxx9}
{{\hat{\nabla}}}_i Z_j = \bigg( -{3 \over \ell}+ h^m Z_m \bigg) \delta_{ij} +{3 \over \ell} Z_i Z_j - Z_i h_j
-{1 \over 2} \Phi (\star_3 Z)_{ij}
\end{eqnarray}
and hence, in particular,
\begin{eqnarray}
\star_3 dZ = - \ell \Phi ({1 \over 2}h +{1 \over \sqrt{3}} \star_3 dB) -{1 \over \sqrt{3}} \ell i_h dB \ .
\end{eqnarray}
Then, on taking the exterior derivative of ({\ref{auxx8}}), and making use of the
gauge field equation ({\ref{geq1}}), one finds the condition
\begin{eqnarray}
\label{auxx10}
\star_3 dh = d \Phi -2 \Phi h - 2 \sqrt{3} \Phi \star_3 dB \ .
\end{eqnarray}
Moreover, on substituting ({\ref{auxx10}}) into ({\ref{geq2}}), one obtains
\begin{eqnarray}
\label{auxx11}
{{\hat{\nabla}}}^i {{\hat{\nabla}}}_i \Phi + \big(-2 \sqrt{3} \star_3 dB -2 h\big)^i {{\hat{\nabla}}} _i \Phi
+ \Phi \bigg( {8 \over \sqrt{3}} h^i \star_3 dB_i +{16 \over 3} \star_3 dB_i \star_3 dB^i
+{8 \over \ell^2} \bigg) =0 \ .
\nonumber \\
\end{eqnarray}

We remark that the conditions ({\ref{ident1}}), together with
({\ref{zdef}}), ({\ref{auxx8}}), ({\ref{auxx8b}}), ({\ref{auxx9}}),  ({\ref{auxx10}}),
({\ref{auxx11}}) and the expression for the Ricci scalar given in ({\ref{eij}}),
are equivalent to the conditions previously obtained on ${\cal{S}}$ when
one identifies the Killing vector generated by the Killing spinor $\epsilon$
 with the Killing vector ${\partial \over \partial u}$. Here, we have not made this identification.
 Nevertheless, we  obtain the same conditions as a consequence of the compactness of
 ${\cal{S}}$.
 
 \subsection{$AdS_5$ Solutions with $F=0$}

It is useful to briefly revisit the
special case of $AdS_5$ with $F=0$, previously considered in \cite{reallads}.
Assuming that there is a way to write
$AdS_5$ in Gaussian null co-ordinates
with regular near-horizon data on a compact horizon section 
${\cal{S}}$, supersymmetry implies that{\footnote{The conditions ${\cal{S}}=H^3$ and $dh=0$ follow directly from considering the integrability
conditions of ({\ref{gv1}}) and ({\ref{gv2}}) with $\Phi=dB=0$.
({\ref{auxx5}}) then implies that $h^2={4 \over \ell^2}$ and
({\ref{auxx4}}) implies $\Delta=0$. We have made
use of compactness, as a maximum principle has been used to establish
${\hat{\nabla}}_i \langle \phi_+, \phi_+ \rangle =0$.}}
${\cal{S}}=H^3$, $\Delta=dh=0$.
However, this is a contradiction
as either ${\cal{S}}$ is non-compact, or if one makes identifications then either the data are not smooth, or there
is a boundary.

As a consequence, $AdS_5$ cannot be written in this fashion,
such that our assumptions about the smoothness and compactness of
${\cal{S}}$ are simultaneously satisfied.

 \section{Lichnerowicz type identities}

A key step in counting the number of supersymmetries preserved by the near horizon geometries of minimal gauged supergravity is to identify the Killing spinors
of  (\ref{ggv1}) with the zero modes of the associated horizon Dirac equations. The Killing spinors are parallel Dirac $Spin_c(3)$ spinors on ${\cal S}$ and so
are zero modes of the associated Dirac equations. The main objective is to establish the converse. Such a result arises from a Lichnerowicz type theorem.

The classical Lichnerowicz theorem states that on any spin closed manifold $M$,
\begin{eqnarray}
\int_M \langle \Gamma^i \nabla_i \epsilon , \Gamma^j \nabla_j \epsilon \rangle
= \int_M \langle \nabla_i \epsilon, \nabla^i \epsilon \rangle + \int_M {R \over 4}
\langle \epsilon , \epsilon \rangle~,
\end{eqnarray}
where here $\nabla$ is the Levi-Civita connection, and $R$ is the Ricci scalar of $M$.
So if $R=0$ it follows that if $\epsilon$ is a zero mode of the Dirac equation, then $\epsilon$
is parallel with respect to the Levi-Civita connection. A similar theorem has been demonstrated  for near horizon geometries
in 11-dimensional supergravity \cite{d11nh}.

To begin, first recall that the KSEs (\ref{ggv1}) are
\begin{eqnarray}
\nabla_i^\pm \eta_\pm=0~.
\label{ggv2}
\end{eqnarray}
The associated horizon Dirac equations are
\begin{eqnarray}
\label{diracpm}
{\cal D}^\pm\eta_\pm\equiv \Gamma^i {{\hat{\nabla}}}_i \eta_\pm + \Psi^\pm \eta_\pm\ ,
\end{eqnarray}
where
\begin{eqnarray}
\Psi^\pm = \mp ({1 \over 4} h_i \Gamma^i -{\sqrt{3} \over 2} \star_3 dB_i \Gamma^i) + 3\big( \mp {i \over 4} \Phi +{1 \over 2 \ell} \big)
+{\sqrt{3} i \over \ell} B_i \Gamma^i~.
\end{eqnarray}
Clearly if $\eta_\pm$ satisfy (\ref{ggv2}), then they are  zero modes of the horizon Dirac equations, i.e.
\begin{eqnarray}
{\cal D}^\pm\eta_\pm=0~.
\end{eqnarray}
To prove the converse, we define
\begin{eqnarray}
I^\pm  &=& \int_{\cal{S}} \Big(\langle \nabla^\pm_i \eta_\pm  , \nabla^{\pm i} \eta_\pm\rangle
 -\langle {\cal D}^\pm\eta_\pm , {\cal D}^\pm\eta_\pm \rangle\Big) \ .
\end{eqnarray}
In order to compute $I^\pm$ it is useful to split $I^\pm$ into three terms. First, note that
\begin{eqnarray}
\label{term1}
\bigg( (\Psi^\pm_i)^\dagger \Psi^{\pm i} - \Psi^\dagger \Psi \bigg) \eta_\pm
&=& \bigg( \pm \big({\sqrt{3} \over \ell} \Phi B_i \Gamma^i +{1 \over 2 \ell} h_i \Gamma^i
-{\sqrt{3} \over \ell} \star_3 dB_i \Gamma^i \big)
\nonumber \\
&-&{1 \over \ell} B^i dB_{ij} \Gamma^j -{\sqrt{3} \over 2 \ell} B_i h_j \epsilon^{ijk} \Gamma_k
+{1 \over \sqrt{3}} h^i \star_3 dB_i
\nonumber \\
&-&{3 \over 8} \Phi^2 -{3 \over 2 \ell^2} \bigg) \eta_\pm \ .
\nonumber \\
\end{eqnarray}
Also
\begin{eqnarray}
\label{term2}
\int_{\cal{S}} \langle {{\hat{\nabla}}}_i \eta_\pm , {{\hat{\nabla}}}^i \eta_\pm \rangle - \langle \Gamma^i {{\hat{\nabla}}}_i \eta_\pm , \Gamma^j {{\hat{\nabla}}}_j \eta_\pm \rangle
&=& \int_{\cal{S}} - {{\hat{\nabla}}}_i \langle \eta_\pm , \Gamma^{ij} {{\hat{\nabla}}}_j \eta_\pm \rangle + \int_{\cal{S}} \langle \eta_\pm , \Gamma^{ij} {{\hat{\nabla}}}_i
{{\hat{\nabla}}}_j \eta_\pm \rangle
\nonumber \\
&=&  \int_{\cal{S}} - {{\hat{\nabla}}}_i \langle \eta_\pm , \Gamma^{ij} {{\hat{\nabla}}}_j \eta_\pm \rangle
\nonumber \\
&+& \int_{\cal{S}} -{1 \over 4} \bigg( -{{\hat{\nabla}}}^i h_i +{1 \over 2} h^2
+dB_{mn} dB^{mn}
\nonumber \\
&+&{3 \over 2} \Phi^2 -{12 \over \ell^2} \bigg)
\langle \eta_\pm , \eta_\pm \rangle \ .
\end{eqnarray}
Note that in the above expression the first term on the RHS is a surface term, this has been retained
because the expression being differentiated in this term
is not $U(1)$ gauge-invariant. Furthermore, we have used the Einstein equations ({\ref{eij}}) in order to compute the
Ricci scalar of ${\cal{S}}$.

The remaining term contributing to $I^\pm$ is
\begin{eqnarray}
\label{term3a}
\int_{\cal{S}} \langle {{\hat{\nabla}}}_i \eta_\pm , \Psi^{\pm i} \eta_\pm \rangle
+ \langle \Psi^\pm_i \eta_\pm , {{\hat{\nabla}}}^i \eta_\pm \rangle
- \langle \Gamma^i {{\hat{\nabla}}}_i \eta_\pm, \Psi^\pm \eta_\pm \rangle - \langle \Psi^\pm \eta_\pm , \Gamma^i {{\hat{\nabla}}}_i \eta_\pm \rangle \ .
\end{eqnarray}
On performing a partial integration, one finds that this expression can be rewritten as
\begin{eqnarray}
\label{term3b}
\int_{\cal{S}}
\langle \eta_\pm, {{\hat{\nabla}}}_i (\Gamma^i \Psi^\pm - \Psi^{\pm i}) \eta_\pm \rangle
+ \langle \eta_\pm , \big( (\Psi^{\pm i} - \Gamma^i \Psi^\pm)^\dagger -  (\Psi^{\pm i} - \Gamma^i \Psi^\pm) \big)
{{\hat{\nabla}}}_i \eta_\pm \rangle
\nonumber \\
+ \int_{\cal{S}} {{\hat{\nabla}}}_i \bigg( \langle \eta_\pm , (\Psi^{\pm i} - \Gamma^i \Psi^\pm) \eta_\pm  \rangle \bigg) \ .
\end{eqnarray}
Again, a surface term has been retained, as the term in the parentheses in the second line is not $U(1)$ gauge invariant.
Note in particular that
\begin{eqnarray}
 \int_{\cal{S}} {{\hat{\nabla}}}_i \bigg( \langle \eta_\pm , (\Psi^{\pm i} - \Gamma^i \Psi^\pm) \eta_\pm \rangle \bigg)
= \int_{\cal{S}}  {{\hat{\nabla}}}_i \bigg( \langle \eta_\pm ,-{\sqrt{3} i \over \ell} B_j \Gamma^{ij} \eta_\pm \rangle \bigg) \ .
\end{eqnarray}
In order to compute the remainder of ({\ref{term3b}}) note that
\begin{eqnarray}
{\rm{Re}} \bigg( \langle \eta_\pm, {{\hat{\nabla}}}_i (\Gamma^i \Psi^\pm - \Psi^{\pm i}) \eta_\pm \rangle  \bigg) = \pm {\sqrt{3} \over \ell}
\langle \eta_\pm , \star_3 dB_i \Gamma^i \eta_\pm \rangle \ .
\end{eqnarray}
In addition,
\begin{eqnarray}
\bigg( \big(\Psi^\pm_i - \Gamma_i \Psi^\pm \big) -  \big(\Psi^\pm_i - \Gamma_i \Psi^\pm \big)^\dagger \bigg) {{\hat{\nabla}}}^i \eta_\pm
= \bigg(-{i \over 2} h_j \epsilon_i{}^{jk} \Gamma_k -{i \over \sqrt{3}} dB_{ij} \Gamma^j \pm i \Phi \Gamma_i \bigg) {{\hat{\nabla}}}^i \eta_\pm \ .
\nonumber \\
\end{eqnarray}
The contribution to $I^\pm$ obtained from this term is evaluated using the following identity
then
\begin{eqnarray}
{\rm Re} \langle \eta_\pm , i X_{ij} \Gamma^j {{\hat{\nabla}}}^i \eta_\pm \rangle
= -{1 \over 2} {\rm Re} \langle \eta_\pm, i X_{ij} \Gamma^{ij} \Gamma^k {{\hat{\nabla}}}_k \eta_\pm \rangle
\pm {1 \over 2} \star_3 X^i {{\hat{\nabla}}}_i \langle \eta_\pm , \eta_\pm \rangle \ ,
\end{eqnarray}
where  $X$ is any real 2-form on ${\cal{S}}$.
Setting $X= -{1 \over 2} \star_3 h +{1 \over \sqrt{3}} dB$, one finds that ({\ref{term3b}}) can be rewritten as
\begin{eqnarray}
\label{term3c}
 \int_{\cal{S}}  {{\hat{\nabla}}}_i \bigg( \langle \eta_\pm ,-{\sqrt{3} i \over \ell} B_j \Gamma^{ij} \eta_\pm \rangle \bigg)
\pm {\sqrt{3} \over \ell} \langle \eta_\pm, \star_3 dB_i \Gamma^i \eta_\pm \rangle \mp {1 \over 4}
h^i {{\hat{\nabla}}}_i \langle \eta_\pm , \eta_\pm \rangle
\nonumber \\
\pm {\rm Re \ } \bigg( \langle \eta_\pm, \big(-i \Phi +{1 \over 2} h_j \Gamma^j -{1 \over \sqrt{3}} \star_3 dB_j \Gamma^j \big)
\Gamma^i {{\hat{\nabla}}}_i \eta_\pm \rangle \bigg) \ .
\end{eqnarray}
To compute $I^\pm$, we take the sum of ({\ref{term1}}), ({\ref{term2}}) and ({\ref{term3c}}), and observe that
the sum of the two surface terms in ({\ref{term2}}) and ({\ref{term3b}}) vanishes. Next, rewrite the $\Gamma^i {{\hat{\nabla}}}_i \eta_\pm$
term arising in ({\ref{term3c}}) in terms of the Dirac operator $\Gamma^i {{\hat{\nabla}}}_i \eta_\pm + \Psi^\pm \eta_\pm$
and  $\Psi^\pm \eta_\pm$.
One then obtains
\begin{eqnarray}
\label{term4}
I^\pm &=& {\rm Re \ } \bigg( \int_{\cal{S}} \langle \eta_\pm , \big( {1 \over \ell} \pm (-i \Phi +{1 \over 2} h_j \Gamma^j -{1 \over \sqrt{3}} \star_3 dB_j \Gamma^j)
 \big(\Gamma^i {{\hat{\nabla}}}_i \eta_\pm + \Psi^\pm \eta_\pm\big) \rangle \bigg)
\nonumber \\
&+&  \int_{\cal{S}} {1 \over 4} {{\hat{\nabla}}}^i h_i \langle \eta_\pm , \eta_\pm \rangle \mp {1 \over 4} h^i {{\hat{\nabla}}}_i \langle \eta_\pm , \eta_\pm \rangle \ .
\end{eqnarray}
We remark that to establish the above identity the only bosonic field equation which was utilized in the above analysis was the trace of
({\ref{eij}}) and was used to evaluate the Ricci scalar of ${\cal{S}}$. The relations ({\ref{ident1}}) and ({\ref{ident2}}) amongst the fields
were not used as they follow from the KSEs.

It is straightforward to observe that for the zero modes of ${\cal D}^\pm$ to be parallel with respect to $\nabla^\pm$ and so Killing spinors, the integrals $I^\pm$ must vanish.  It is clear that if ${\cal D}^-\eta_-=0$, then
$I^-=0$ and so $\nabla^-\eta_-=0$. All the zero modes of the horizon Dirac operator ${\cal D}^-$ are Killing spinors.  Next let us turn to $I^+$. If ${\cal D}^+\eta_+=0$, then $I^+$ does not vanish
unless one imposes the condition $\langle \eta_+, \eta_+ \rangle = \mathrm {const}$.  Thus we have established the following two statements
\begin{eqnarray}
\label{dr1}
\nabla_i^+\eta_+=0 \Longleftrightarrow {\cal D}^+\eta_+ =0 ~~ \mathrm{and} ~~ \langle \eta_+, \eta_+ \rangle = \mathrm{const}~,
\end{eqnarray}
and
\begin{eqnarray}
\label{dr2}
\nabla_i^-\eta_-=0 \Longleftrightarrow  {\cal D}^-\eta_- =0  \ .
\end{eqnarray}
This concludes the proof of the Lichnerowicz type theorems for the horizons of minimal gauged supergravity.

\section{Supersymmetry of near horizon geometries}

Before we proceed to identify the number of supersymmetries preserved by the near horizon geometries of minimal gauged supergravity, we
shall first examine in more detail the spinors on ${\cal S}$.  The spinors that enter into the KSEs of minimal gauged 5-dimensional supergravity (\ref{grav})
are Dirac $Spin_c(4,1)=Spin(4,1)\cdot U(1)$ spinors and so sections of the bundle\footnote{Typically in $Spin_c$ structures the bundles $S$ and $L$ may not be well-defined
but their product is. We shall not expand on this and we shall assume that both $S$ and $L$ are well-defined bundles.} $S\otimes L$, where $S$ is the spin bundle and $L$ is a $U(1)$ bundle on the spacetime. When these are restricted
on ${\cal S}$,  $S\otimes L$ decomposes as $S\otimes L=S^+\otimes L\oplus S^-\otimes L$, where the signs are referred to the projections $\Gamma_\pm$
and $S^\pm\otimes L$ are $Spin_c(3)$ bundles. We have identified $L$ with its restriction on ${\cal S}$.  Furthermore, the horizon Dirac operators act as
${\cal D}^\pm: \Gamma(S^\pm \otimes L)\rightarrow \Gamma(S^\pm \otimes L)$, where   $\Gamma(S^\pm \otimes L)$ are the smooth sections of $S^\pm \otimes L$.

Next let us return to examine the number of supersymmetries preserved by the near horizon geometries. We have established that if a near horizon geometry is
supersymmetric,  there must exist at least one  non-vanishing spinor $\eta_+$ such that $\nabla^+_i\eta_+=0$.
Since on ${\cal S}$ there can be up to two linearly independent $\nabla^+$-parallel spinors,  there are two cases to investigate. First suppose that there are strictly two $\nabla^+$-parallel spinors. In this case,  one can show that
the near horizon geometry is $AdS_5$ with $F=0$. This follows directly from the algebraic condition ({\ref{alg1}}), together with the conditions on the geometry
 in section 4. Indeed
note that on expanding out ({\ref{alg1}}), the vanishing of the term zeroth order in gamma matrices
implies that
\begin{eqnarray}
\Phi=0
\end{eqnarray}
and hence
\begin{eqnarray}
\Delta=0
\end{eqnarray}
as a consequence ({\ref{ident1}}). Then ({\ref{auxx10}}) implies that
\begin{eqnarray}
dh=0 \ .
\end{eqnarray}
Returning to ({\ref{alg1}}), the remaining conditions imply that
\begin{eqnarray}
dB=0 \ .
\end{eqnarray}
Hence $F=0$, and it is straightforward to show that the remaining conditions on the geometry
listed in the previous section imply that the solution is $AdS_5$.

It remains to investigate the horizons which admit   strictly one   linearly independent $\nabla^+$-parallel spinor.  In such a case,  we have\footnote{${\cal D}^+$ may have more than
   one zero mode as it is not a priori necessary for all zero modes to satisfy the normalization condition $\langle \eta_+, \eta_+ \rangle =1$.}   $\mathrm{dim}_{{\mathbb{C}}}\, \mathrm{Ker}\, {\cal D}^+\geq 1$. To proceed, we shall demonstrate that
\begin{eqnarray}
\mathrm{dim}_{{\mathbb{C}}}\, \mathrm{Ker}\, {\cal D}^+=\mathrm{dim}_{{\mathbb{C}}}\, \mathrm{Ker}\, {\cal D}^-~.
\label{kerker}
\end{eqnarray}
To see this, first observe that the adjoint of ${\cal D}^+$,  $({\cal D}^+)^\dagger: \Gamma(S^+ \otimes L)\rightarrow \Gamma(S^+ \otimes  L)$. Since ${\cal D}^+$
has the same principal symbol as a $U(1)$ twisted Dirac operator, and as it is defined on the odd dimensional manifold ${\cal S}$, the index vanishes, $\mathrm{Index} ({\cal D}^+)=
\mathrm{dim}_{{\mathbb{C}}}\, \mathrm{Ker}\, {\cal D}^+- \mathrm{dim}_{{\mathbb{C}}}\, \mathrm{Ker}\,({\cal D}^+)^\dagger=0$.  As a result, we obtain
\begin{eqnarray}
\mathrm{dim}_{{\mathbb{C}}}\, \mathrm{Ker}\, {\cal D}^+=\mathrm{dim}_{{\mathbb{C}}}\, \mathrm{Ker}\,({\cal D}^+)^\dagger~.
\end{eqnarray}
It remains to relate the kernels of $({\cal D}^+)^\dagger$ and ${\cal D}^-$. First observe that
\begin{eqnarray}
({\cal{D}}^+)^\dagger = -\Gamma^i {{\hat{\nabla}}}_i -({1 \over 4} h_i \Gamma^i - {\sqrt{3} \over 2} \star_3 dB_i
\Gamma^i)+3({i \over 4} \Phi +{1 \over 2 \ell}) -{\sqrt{3}i \over \ell} B_i \Gamma^i \ .
\end{eqnarray}
Next set
\begin{eqnarray}
\label{chirchange}
\eta_- = \Gamma_- \eta'_+ \ ,
\end{eqnarray}
which induces an isomorphism between $\Gamma(S^+\otimes L)$ and $\Gamma(S^-\otimes L)$, and observe that
\begin{eqnarray}
{\cal D}^-\eta_-=\Gamma_-({\cal{D}}^+)^\dagger \eta'_+
\end{eqnarray}
which establishes (\ref{kerker}).

The Lichnerowicz type theorem we have shown for the ${\cal D}^-$ horizon Dirac equation ({\ref{dr2}}) identifies the $\nabla^-$-parallel spinors with
the zero modes of  ${\cal D}^-$.   First suppose that  $\mathrm{dim}_{{\mathbb{C}}}\, \mathrm{Ker}\, {\cal D}^- = 2$. In such a case,
there are two $\nabla^-$-parallel spinors and so the near horizon geometries preserves
3/4 of the supersymmetry. It has been shown that all such solutions are locally isometric to
$AdS_5$ with vanishing flux $F=0$, \cite{preon1, preon2}. The remaining case is $\mathrm{dim}_{{\mathbb{C}}}\, \mathrm{Ker}\, {\cal D}^- =1$. In this case, the horizons preserve
$1/2$ of the supersymmetry.

To summarize, we have demonstrated that the near horizon geometries of minimal gauged 5-dimensional supergravity preserve at least half of the supersymmetry. If they preserve
 a larger fraction of supersymmetry, then they are locally
isometric to $AdS_5$ and $F=0$.

\section{The ${\mathfrak{sl}}(2,{\mathbb{R}})$ symmetry}

In this section, we prove that the half-supersymmetric near-horizon geometries admit a 
${\mathfrak{sl}}(2,{\mathbb{R}})$ symmetry. The analysis closely follows that performed for M-horizons in
\cite{mhor}. To proceed, we first note that the most general Killing spinor
can be written as
\begin{eqnarray}
\epsilon = \eta_+ + u \Gamma_+ \Theta_- \eta_- + \eta_- + r \Gamma_- \Theta_+ \eta_+
+ ur \Gamma_- \Theta_+ \Gamma_+ \Theta_- \eta_-
\end{eqnarray}
where 
\begin{eqnarray}
\Theta_\pm = \big({1 \over 4} h \pm {1 \over 2 \sqrt{3}} \star_3 dB\big)_i
\Gamma^i \mp{i \over 2} \Phi -{1 \over 2 \ell} \ .
\end{eqnarray}
Hence, for exactly $1/2$ supersymmetric solutions, we can without loss of generality take
the two linearly independent Killing spinors to be
\begin{eqnarray}
\epsilon_1 = \eta_- + u \eta_+ + ru \Gamma_- \Theta_+ \eta_+, \qquad
\epsilon_2 = \eta_+ + r \Gamma_- \Theta_+ \eta_+
\end{eqnarray}
where
\begin{eqnarray}
\eta_+ = \Gamma_+ \Theta_- \eta_- \ .
\end{eqnarray}
The condition ({\ref{auxx7}}) is equivalent to
\begin{eqnarray}
\Theta_+ \eta_+ = -{i \over 2}\Phi \eta_+ \ .
\end{eqnarray}
The two Killing spinors $\epsilon_1, \epsilon_2$ can then be further simplified to
\begin{eqnarray}
\epsilon_1 = \eta_- +u \eta_+ -{i \over 2} ur \Phi \Gamma_- \eta_+, \qquad
\epsilon_2 = \eta_+ -{i \over 2}r \Phi \Gamma_- \eta_+ \ .
\end{eqnarray}
Next we define three 1-form spinor bilinears $K_1, K_2, K_3$ by
\begin{eqnarray}
K_1 &=& {\rm Re} \ \bigg( {\cal{B}} \big(\epsilon_1, \Gamma_\mu \epsilon_2\big) \bigg)
{\bf{e}}^\mu
\nonumber \\
K_2 &=& {\cal{B}} \big(\epsilon_2, \Gamma_\mu \epsilon_2\big) {\bf{e}}^\mu
\nonumber \\
K_3 &=& {\cal{B}} \big(\epsilon_1, \Gamma_\mu \epsilon_1\big) {\bf{e}}^\mu \ ,
\end{eqnarray}
where ${\cal{B}}$ is the $Spin(4,1) \times U(1)$ invariant inner product defined in
({\ref{spip}}). Then from the analysis in Appendix B, it follows that $K_1, K_2, K_3$ are
associated with vector fields which are isometries that also preserve $F$. 

We proceed to compute the components of $K_1, K_2, K_3$: one obtains
\begin{eqnarray}
\label{sl21}
K_1 &=& \bigg(r^2 u \Delta \parallel \eta_+ \parallel^2 +r \Delta \parallel \eta_-
\parallel^2 \bigg) {\bf{e}}^+ -2u \parallel \eta_+ \parallel^2 {\bf{e}}^-
+ V_i {\bf{e}}^i
\nonumber \\
K_2 &=& r^2 \Delta \parallel \eta_+ \parallel^2 {\bf{e}}^+
-2 \parallel \eta_+ \parallel^2 {\bf{e}}^-
\nonumber \\
K_3 &=& \bigg(2 \parallel \eta_- \parallel^2 +2 ru \Delta \parallel \eta_- \parallel^2
+r^2 u^2 \Delta \parallel \eta_+ \parallel^2 \bigg) {\bf{e}}^+
-2 u^2 \parallel \eta_+ \parallel^2 {\bf{e}}^- +2u V_i {\bf{e}}^i \ ,
\nonumber \\
\end{eqnarray}
where
\begin{eqnarray}
V_i = {\rm Re} \ \langle \Gamma_+ \eta_- , \Gamma_i \eta_+ \rangle  \ .
\end{eqnarray}
We remark that in order to obtain ({\ref{sl21}}) we have made use of the identities
\begin{eqnarray}
{\rm Re} \ \langle \Gamma_+ \eta_- , -i \Phi \eta_+ \rangle = \Delta \parallel \eta_-
\parallel^2 \ ,
\end{eqnarray}
\begin{eqnarray}
\langle \Gamma_- \eta_+, \eta_- \rangle - \langle \eta_-, \Gamma_- \eta_+ \rangle
=-2i \Phi \parallel \eta_- \parallel^2 \ .
\end{eqnarray}
The vector fields dual to the 1-forms in ({\ref{sl21}})
are 
\begin{eqnarray}
\label{sl22}
K_1 &=& -2u \parallel \eta_+ \parallel^2 \partial_u +2r \parallel \eta_+ \parallel^2 \partial_r + V^i {\tilde{\partial}}_i
\nonumber \\
K_2 &=& -2 \parallel \eta_+ \parallel^2 \partial_u
\nonumber \\
K_3 &=& -2u^2 \parallel \eta_+ \parallel^2 \partial_u + \bigg(2 \parallel \eta_-
\parallel^2 +4ru \parallel \eta_+ \parallel^2 \bigg) \partial_r +2u V^i {\tilde{\partial}}_i
\end{eqnarray}
where to obtain these vector fields we have made use of the condition
\begin{eqnarray}
\label{nm2}
h^i V_i = \Delta \parallel \eta_- \parallel^2 -2 \parallel \eta_+ \parallel^2 \ .
\end{eqnarray}
The vector fields listed in ({\ref{sl22}}) are (formally) identical to
those obtained from the corresponding 11-dimensional calculation given in
\cite{mhor}.
In particular, on imposing the Killing condition ${\cal{L}}_{K_a} g=0$, one finds 
that
\begin{eqnarray}
{\hat{\nabla}}_{(i}V_{j)}=0, \qquad {\cal{L}}_V \Delta=0, \qquad {\cal{L}}_V h=0
\end{eqnarray}
together with the condition
\begin{eqnarray}
\label{nm1}
d \parallel \eta_- \parallel^2 + V + \parallel \eta_- \parallel^2 h =0 \ .
\end{eqnarray}
This further implies that
\begin{eqnarray}
{\cal{L}}_V \parallel \eta_- \parallel^2 =0 \ .
\end{eqnarray}

Making use of these conditions, it is straightforward to show that the
vector fields given in ({\ref{sl22}}) satisfy 
\begin{eqnarray}
[K_1, K_2] = 2 \parallel \eta_+ \parallel^2 K_2,
\quad [K_2, K_3]= -4 \parallel \eta_+ \parallel^2, \quad [K_3, K_1]=2 \parallel \eta_+ \parallel^2 K_3 \ .
\end{eqnarray}
So the half-supersymmetric near-horizon geometries admit a ${\mathfrak{sl}}(2,\mathbb{R})$ symmetry subalgebra. 

\subsection{Solutions with $V=0$}

In the special case $V=0$, the conditions ({\ref{nm1}}) and ({\ref{nm2}}) imply that
\begin{eqnarray}
d \Delta - \Delta h=0 \ .
\end{eqnarray}
with $\Delta \neq 0$. Hence $d h=0$. 
The $++$ component of the Einstein equations then implies that
\begin{eqnarray}
\Delta h^2 =0 \ .
\end{eqnarray}
As $\Delta \neq 0$, we obtain $h=0$ and so $\Delta$ is constant. 
However, the $+-$ component of the Einstein equations then implies that
\begin{eqnarray}
{1 \over 3} (dB)_{mn} (dB)^{mn}+{4 \over \ell^2}=0 \ .
\end{eqnarray}
Hence, there are no solutions with $V=0$.

\section{Conclusions}

We have proven that the near-horizon geometries of minimal 5-dimensional gauged supergravity preserve
at least half of the supersymmetry. Moreover, if near-horizon geometries preserve a larger fraction, then they are locally isometric to $AdS_5$ and the 2-form field strength $F$ vanishes. An application of these
results is that all half-supersymmetric 5-dimensional horizons admit an $\mathfrak{sl}(2, {\mathbb{R}})$ symmetry algebra 
which follows from a similar argument to that in \cite{mhor}.

The proof of this result utilizes in an essential way the compactness of the horizon sections. The analysis proceeds by first identifying the independent field and Killing spinor equations
after integrating the latter along the lighcone directions.
Next, we have used these  to relate the Killing spinors of the near horizon  to the zero modes of two  Dirac operators defined on horizon sections, which is done
by proving two Lichnerowicz type theorems. To establish our result, one has to count the number of  zero modes of the horizon Dirac operators,  which in turn counts  the number of supersymmetries
preserved by near horizon geometries. For this, we have used the vanishing of the index of
one of the two horizon Dirac operators.

Although many steps in the proof appear to depend on the details of minimal 5-dimensional gauged supergravity, like its field and KSEs, this may not be the case.
It is likely that near horizon geometries of odd-dimensional supergravities, without higher curvature corrections,
generically preserve at least two supersymmetries. This is in agreement with the recently
established property
that M-horizons  preserve an even number of supersymmetries \cite{mhor}. It is also likely that our results extend to the near horizon geometries of even-dimensional supergravities.
However, there are some  differences. For example, the index of the horizon Dirac operators is not expected to vanish. This may lead to
an expression for the number of supersymmetries preserved in terms of the index of a Dirac  operator. In turn, this will relate the number of supersymmetries preserved by a near horizon geometry
to the topology of the horizon sections.

\acknowledgments

\noindent  J. Gutowski is supported by the STFC grant, ST/1004874/1. G. Papadopoulos  is partially supported by  the STFC rolling grant ST/G000/395/1. J. Grover is supported by FCT-Portugal, under the grant SFRH/BPD/78142/2011, and the project PTDC/FIS/116625/2010, as well as by the NRHEPÐ295189  FP7-PEOPLE-2011-IRSES Grant.
The authors would like to thank J. Figueroa-O'Farrill, J.  Lotay and M. Singer for useful discussions.

\appendix

\section{Spin Connection and Curvature}

The non-vanishing components of the spin connection in
the frame basis ({\ref{nhbasis}}) are
\begin{eqnarray}
&&\Omega_{-,+i} = -{1 \over 2} h_i~,~~~
\Omega_{+,+-} = -r \Delta, \quad \Omega_{+,+i} ={1 \over 2} r^2(  \Delta h_i - \partial_i \Delta),
\cr
&&\Omega_{+,-i} = -{1 \over 2} h_i, \quad \Omega_{+,ij} = -{1 \over 2} r dh_{ij}~,~~~
\Omega_{i,+-} = {1 \over 2} h_i, \quad \Omega_{i,+j} = -{1 \over 2} r dh_{ij},
\cr
&&\Omega_{i,jk}= \hat\Omega_{i,jk}~,
\end{eqnarray}
where $\hat\Omega$ denotes the spin-connection of the 3-manifold ${{\cal{S}}}$ with basis ${\bf{e}}^i$.
If $f$ is any function of spacetime, then frame derivatives are expressed in terms of co-ordinate derivatives  as
\begin{eqnarray}
\partial_+ f &=& \partial_u f +{1 \over 2} r^2 \Delta \partial_r f~,~~
\partial_- f = \partial_r f~,~~
\partial_i f = {\tilde{\partial}}_i f -r \partial_r f h_i \ .
\end{eqnarray}
The non-vanishing components of the Ricci tensor is the
 basis ({\ref{nhbasis}}) are
\begin{eqnarray}
R_{+-} &=& {1 \over 2} {{\hat{\nabla}}}^i h_i - \Delta -{1 \over 2} h^2~,~~~
R_{ij} = {\hat{R}}_{ij} + {{\hat{\nabla}}}_{(i} h_{j)} -{1 \over 2} h_i h_j
\nonumber \\
R_{++} &=& r^2 \big( {1 \over 2} {{\hat{\nabla}}}^2 \Delta -{3 \over 2} h^i {{\hat{\nabla}}}_i \Delta -{1 \over 2} \Delta {{\hat{\nabla}}}^i h_i + \Delta h^2
+{1 \over 4} (dh)_{ij} (dh)^{ij} \big)
\nonumber \\
R_{+i} &=& r \big( {1 \over 2} {{\hat{\nabla}}}^j (dh)_{ij} - (dh)_{ij} h^j - {{\hat{\nabla}}}_i \Delta + \Delta h_i \big) \ ,
\end{eqnarray}
where ${\hat{R}}$ is the Ricci tensor of the horizon section ${\cal S}$ in the ${\bf{e}}^i$ frame.

\section{Supersymmetry Conventions}

We first present a matrix representation of $\mathrm{Cliff}(4,1)$ adapted to the basis ({\ref{nhbasis}}).
The space of Dirac spinors is identified with ${\mathbb{C}}^4$ and we set
\begin{eqnarray}
\Gamma_i = \begin{pmatrix}  \sigma^i \ \ \ \ \  0 \cr   \ \ 0 \ \ -\sigma^i \end{pmatrix}, \qquad
\Gamma_- = \begin{pmatrix} \ \ 0  \ \ \ \sqrt{2}\, {\mathbb{I}}_2 \cr  0 \ \ \ \ \ 0 \end{pmatrix}, \qquad
\Gamma_+ = \begin{pmatrix}  \ \ 0 \ \ \ \ \ 0 \cr \sqrt{2}\, {\mathbb{I}}_2 \ \ \ \ 0 \end{pmatrix}
\end{eqnarray}
where $\sigma^i$, $i=1,2,3$ are the Hermitian Pauli matrices $\sigma^i \sigma^j = \delta^{ij} {\mathbb{I}}_2 + i \epsilon^{ijk} \sigma^k$.
Note that
\begin{eqnarray}
\Gamma_{+-} = \begin{pmatrix} -{\mathbb{I}}_2 \ \ \ \ \ 0 \cr \ \ \ 0 \ \ \ {\mathbb{I}}_2 \end{pmatrix}~,
\end{eqnarray}
and hence
\begin{eqnarray}
\Gamma_{+-123} = -i {\mathbb{I}}_4~.
\end{eqnarray}
It will be convenient to decompose the spinors into positive and negative chiralities
with respect to the lightcone directions as
\begin{eqnarray}
\epsilon = \epsilon_+ + \epsilon_-~,
\end{eqnarray}
where
\begin{eqnarray}
\Gamma_{+-} \epsilon_\pm = \pm \epsilon_\pm, \qquad {\rm or \ equivalently} \qquad \Gamma_\pm \epsilon_\pm =0~.
\end{eqnarray}
With these conventions, note that
\begin{eqnarray}
\Gamma_{ij} \epsilon_\pm = \mp i \epsilon_{ij}{}^k \Gamma_k \epsilon_\pm,
\qquad \Gamma_{ijk} \epsilon_\pm = \mp i \epsilon_{ijk} \epsilon_\pm~.
\end{eqnarray}

The Dirac representation of $Spin(4,1)$ decomposes under $Spin(3)=SU(2)$ as ${\mathbb{C}}^4={\mathbb{C}}^2\oplus {\mathbb{C}}^2$ each subspace
specified by the lightcone projections $\Gamma_\pm$. On each ${\mathbb{C}}^2$,
we have made use of the $Spin(3)$-invariant inner product  $\langle , \rangle$ which is identified with the standard Hermitian
inner product. On ${\mathbb{C}}^2\oplus {\mathbb{C}}^2$, the Lie algebra of
$Spin(3)$ is spanned  by $\Gamma_{ij}$, $i,j=1,2,3$. In particular,  note that $(\Gamma_{ij})^\dagger = - \Gamma_{ij}$.

It will also be useful to introduce a non-degenerate $Spin(4,1) \times U(1)$ invariant inner product
${\cal{B}}$ by
\begin{eqnarray}
\label{spip}
{\cal{B}} (\epsilon, \eta) = \langle \big(\Gamma_+-\Gamma_-\big) \epsilon, \eta \rangle \ .
\end{eqnarray}
It is straightforward to show that
\begin{eqnarray}
{\cal{B}} (\epsilon, \Gamma_\mu \eta)+{\cal{B}} (\Gamma_\mu \epsilon,  \eta) &=&0
\nonumber \\
{\cal{B}} (\epsilon, \Gamma_{\mu \nu} \eta)+{\cal{B}} (\Gamma_{\mu \nu} \epsilon,  \eta) &=&0
\nonumber \\
{\cal{B}} (\epsilon, \Gamma_{\mu \nu \rho} \eta)-{\cal{B}} (\Gamma_{\mu \nu \rho}\epsilon,  \eta) &=&0 \ .
\end{eqnarray}
In addition, if $\epsilon_1, \epsilon_2$ are Killing spinors, and if
\begin{eqnarray}
K_\mu = {\cal{B}}(\epsilon_1, \Gamma_\mu \epsilon_2)
\end{eqnarray}
then the KSE ({\ref{grav}}) implies that
\begin{eqnarray}
\nabla_\nu K_\mu = {\cal{B}} \bigg( \epsilon_1, \bigg(-{i \over 2 \sqrt{3}}
F^{\rho \sigma}\Gamma_{\nu \mu \rho \sigma} -{2i \over \sqrt{3}}F_{\nu \mu}
+{1 \over \ell} \Gamma_{\nu \mu} \bigg) \epsilon_2 \bigg) \ .
\end{eqnarray}
Hence
\begin{eqnarray}
\nabla_{(\mu} K_{\nu)}=0 \ .
\end{eqnarray}
Furthermore, one also obtains from ({\ref{grav}})
\begin{eqnarray}
d {\cal{B}}(\epsilon_1, \epsilon_2) = {2i \over \sqrt{3}} i_K F
\end{eqnarray}
and hence
\begin{eqnarray}
{\cal{L}}_K F =0 \ .
\end{eqnarray}
So one obtains isometries, which also preserve $F$, from such 1-form spinor bilinears.

\end{document}